\documentclass[twocolumn,prd,nofootinbib,aps,prl,floats,floatfix,amsmath,amssymb,longbibliography,secnumarabic]{revtex4-1} %
\usepackage[english]{babel}
\usepackage{csquotes}

\usepackage[letterpaper,top=2cm,bottom=2cm,left=3cm,right=3cm,marginparwidth=1.75cm]{geometry}

\usepackage{amsmath}
\usepackage{graphicx,slashed}
\usepackage[colorlinks=true,citecolor=red]{hyperref}

\newcommand{\be}{\begin{equation}}
\newcommand{\ee}{\end{equation}}
\def\sfrac#1#2{{\textstyle{#1\over #2}}}
\newcommand{\bea}{\begin{eqnarray}}
\newcommand{\eea}{\end{eqnarray}}
\newcommand{\nn}{\nonumber}
\newcommand{\E}{{\scriptscriptstyle E}}
\def\sfrac#1#2{{\textstyle{#1\over #2}}}

\begin{document}

\title{Comment on ``Standard Model Mass Spectrum and Interactions In The
Holomorphic Unified Field Theory''}
\author{James M.\ Cline}
\email{jcline@physics.mcgill.ca}
\affiliation{McGill University Department of Physics \& Trottier Space Institute, 3600 Rue University, Montr\'eal, QC, H3A 2T8, Canada}

\begin{abstract}
In \url{https://arxiv.org/abs/2508.02747} a theory that unifies gravity and the Standard Model was proposed.  It is finite, unitary, and accurately predicts all parameters of the
Standard Model in terms of only two input parameters.  Here I point out one prediction that was overlooked, namely that the photon acquires a mass at one loop due to the lack of gauge invariance in the nonlocal interaction vertices of the proposed model.  In a second iteration,
 a red herring concerning Wick rotation is clarified, and the difficulties relating to gauge invariance in the theory are further spelled out. I point out that their model apparently has a domain wall problem.

\end{abstract}

\maketitle
\section{Original comment}

Recently Ref.\ \cite{Moffat:2025awg} 
elaborated on some previous proposals by the same authors featuring a theory of quantum gravity, rendered finite by inserting a nonlocal operator $e^{-\Box/M_*^2}$ into the action.  Here $\Box$ is the 4D d'Alembertian  and $M_*$ is taken to be proportional to a unification scale of order $10^{16}\,$GeV.   Thus factors of $e^{-p^2/M_*^2}$ appear in Wick-rotated loop diagrams and make them convergent.  These are not a regulator in the usual sense, since they are intrinsic to the definition of the theory, and $M_*$ has a direct physical interpretation, rather than being an arbitrary cutoff.

By extending this approach to the Standard Model, and choosing a particular assignment of Froggatt-Nielsen charges to generate Yukawa matrices of a certain texture, the authors were amazingly able to reproduce all the parameters of the standard model to within experimental precision, by varying only two adjustable parameters in the model.

However, the authors did not consider the implications of their model for electroweak precision observables.  Here I take a first step\footnote{and hopefully last} by computing the one-loop correction to the photon mass, which arises since the regulator function is not gauge invariant.  One can set the external photon momentum to zero and compute the vacuum polarization 
(using metric signature $(+,-,-,-)$)
\bea
   \Pi^{\mu\nu}(0) &=&  e^2\int{ d^{\,4}p\over (2\pi)^4}
    {{\rm tr}(\gamma^\mu( \slashed{p}+m)\gamma^\nu
   ( \slashed{p}+m))\over
    (p^2-m^2 + i\epsilon)^2}\nn\\
    &\times& \exp({2p^2/M_*^2})\nn\\
    &\sim& {\alpha\over 4\pi} M_*^2\,\eta^{\mu\nu}
\eea
Therefore, despite the incredible successes of the model in the flavor sector, it is in tension with at least one observable, and probably more, beyond tree level.

\section{Reply to comment on comment}

Recently Ref.\ \cite{Moffat:2025awg} proposed a theory of
everything that is purported to be finite, unitary, and reproduces the Standard Model (SM)  with only two adjustable parameters.
An essential ingredient of the theory is the presence of
nonlocal operators $e^{-\Box/M_*^2}$ in the interaction vertices (where $\Box$ is the 4D d'Alembertian) that make loop diagrams finite without the need for any regulator.
These operators, since they involve ordinary rather than covariant derivatives, are not gauge invariant, and therefore one might be concerned about how such a theory could possibly be consistent.
I raised this concern in a comment \cite{Cline:2025itx}, pointing out that for example, the photon gets a calculable mass at one loop.
The authors have commented on my comment \cite{Moffat:2025smm}
with arguments that could be misconstrued as being reasonable, and hence deserve a reply.

Distracting from the main point, in their comment the authors seem to be confused about Wick rotations.
Here I repeat my computation of the photon vacuum polarization, which originally omitted showing the middle step where the Wick rotation to Euclidean momentum $p_\E$ was performed:
\bea
   \Pi^{\mu\nu}(0) &=&  -ie^2\int{ d^{\,4}p\over (2\pi)^4}
    {{\rm tr}(\gamma^\mu( \slashed{p}+m)\gamma^\nu
   ( \slashed{p}+m))\over
    (p^2-m^2 + i\epsilon)^2}\nn\\
    &\times& \exp({2p^2/M_*^2})\label{mink}\\
    &=& 4e^2\eta^{\mu\nu}\int{ d^{\,4}p_\E\over (2\pi)^4}
    {\sfrac12 p_\E^2 + m^2 \over
    (p_\E^2+m^2)^2}e^{-2p_\E^2/M_*^2}\nn\\
    &\sim& {\alpha\over 4\pi} M_*^2\,\eta^{\mu\nu}
    \label{photon-mass}
\eea
Hence we all agree, the exponential factor makes the integral converge, not diverge, despite superficial appearances in Eq.\ (\ref{mink}).

Let us come back to the issue of gauge symmetry, or lack thereof.  Ref.\ \cite{Moffat:2025awg} claims repeatedly that their construction is gauge-invariant, but gives no hint of how this is supposed to come about despite the regulator being gauge-noninvariant.  The comment on the comment \cite{Moffat:2025smm} however, supplies this information, citing two references \cite{Evens:1990wf, Moffat:1990jj}.  Both of them use the exponential regulator, but the first paper uses it as a regulator in the usual sense, in which the cutoff is to be taken to infinity, while the second paper uses it in the sense appropriate to the present discussion, where the cutoff is regarded as a finite parameter with physical significance.  

This distinction is important.  In fact the authors of
Ref.\ \cite{Evens:1990wf} discuss the problems that arise from 
regarding the nonlocal regulator as having physical meaning, considering as an example a scalar field theory with regulator $\exp[(\Box-m^2)/\Lambda^2]$: keeping $\Lambda$ finite, it leads to
``off-shell noncausality at the perturbative level as well as instability
and a breakdown of the initial-value problem beyond perturbation theory.''  They go on to say
\begin{displayquote}
These problems are fatal to [such a model] as
any sort of fundamental theory, but they pose no obstacle
to regarding it as a perturbative regularization of the local action which results from taking $\Lambda$ to infinity.
\end{displayquote}
Ref.\ \cite{Evens:1990wf} constructs a nonlocal generalization of gauge invariance that reduces to the conventional one when $\Lambda\to\infty$.  It thus seems to be feasible as a regulator for gauge theories, but it is complicated to implement, and hence it did not catch on as a practical tool for perturbative calculations.\footnote{Also, the claim that it circumvented the fermion doubling problem turned out to be mistaken \cite{RPW}.}

\begin{figure*}[t]
\centerline{\includegraphics[width=2\columnwidth]{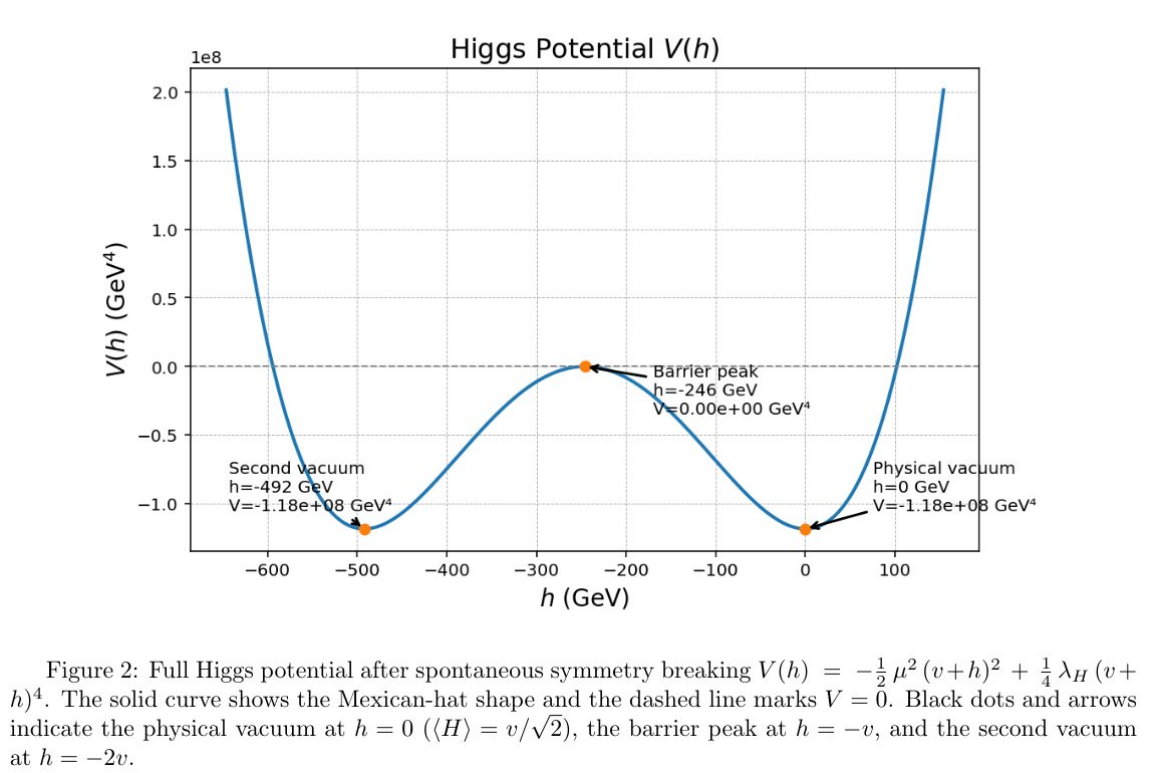}}
\caption{Reproduction of Fig.\ 2 from Ref.\ \cite{Moffat:2025awg}, highlighting its prediction of a second Higgs vacuum.}
\label{fig:1}
\end{figure*}

Unfortunately Ref.\ \cite{Moffat:1990jj}
only again asserts that gauge invariance is maintained, but it gives no hint as to which additional interactions should be added to cancel unwanted results such as Eq.\ (\ref{photon-mass}).  Therefore we are back to Ref.\ \cite{Evens:1990wf}, restricted to finite $\Lambda = M_*$.    The short answer is that one constructs a (finite) counterterm 
Lagrangian, order-by-order in the couplings, to cancel amplitudes that violate the Ward identities,
and thereby restore gauge invariance.  Notice that a similar procedure could be done for any gauge-breaking regulator.

It would seem to be a hard road to follow, in checking whether precision electroweak observables are the same in this theory as in the standard model.
This question in fact was enunciated in Ref.\ \cite{Moffat:1990jj}:
\begin{displayquote}
Hopefully, future high-energy accelerators will be able to check the radiative Higgs corrections predicted by the nonlocal electroweak field theory, so that we can compare them with those predicted by the GSW [Glashow-Weinberg-Salam] theory.
\end{displayquote}
Apparently, 35 years after the theory was first proposed, we are still awaiting the answer to this question, since no mention of the issue was made in Ref.\ \cite{Moffat:2025awg}.

On the other hand, from their description of the Higgs potential, there seems to be already at tree level a dramatic difference between their model and the Standard Model, since they say it is a double-well potential, with a degenerate minimum at $h = -2v$, as shown in Fig.\ \ref{fig:1}.  This is very different from the SM, where the entire Higgs vacuum manifold corresponds to a single state, due to gauge symmetry.  The authors note that ``A symmetric second minimum occurs at $h = -2v$, but once the real-slice vacuum is
fixed at $h = 0$, tunnelling to this well is exponentially suppressed.''  However, the formation of domain walls would not be suppressed during the electroweak phase transition in the early Universe, so perhaps this provides a more tractable way of distinguishing the HUFT  from the SM, compared to the uninviting prospect of trying to predict precision electroweak observables in the HUFT.

\medskip
I thank Richard Woodard for helpful discussions.  He sends his greetings to J.\ Moffat.

\bibliographystyle{utphys}
\bibliography{sample}

\end{document}